# Josephson supercurrent through the topological surface states of strained bulk HgTe

Jeroen B. Oostinga,[1] Luis Maier,[1] Peter Schüffelgen,[1] Daniel Knott,[1] Christopher Ames,[1] Christoph Brüne,[1] Grigory Tkachov,[2] Hartmut Buhmann,[1] and Laurens W. Molenkamp[1]

[1] *Physikalisches Institut (EP3), University of Würzburg, Am Hubland, D-97074 Würzburg, Germany*
[2] *Institut für Theoretische Physik und Astrophysik (TP4), University of Würzburg, Am Hubland, D-97074 Würzburg, Germany*

**Strained bulk HgTe is a three-dimensional topological insulator, whose surface electrons have a high mobility ($\sim$30,000 cm$^2$/Vs), while its bulk is effectively free of mobile charge carriers. These properties enable a study of transport through its unconventional surface states without being hindered by a parallel bulk conductance. Here, we show transport experiments on HgTe-based Josephson junctions to investigate the appearance of the predicted Majorana states at the interface between a topological insulator and a superconductor. Interestingly, we observe a dissipationless supercurrent flow through the topological surface states of HgTe. The current-voltage characteristics are hysteretic at temperatures below 1 K with critical supercurrents of several microamperes. Moreover, we observe a magnetic field induced Fraunhofer pattern of the critical supercurrent, indicating a dominant $2\pi$-periodic Josephson effect in the unconventional surface states. Our results show that strained bulk HgTe is a promising material system to get a better understanding of the Josephson effect in topological surface states, and to search for the manifestation of zero-energy Majorana states in transport experiments.**

A strained, undoped layer of HgTe thicker than $\sim$50 nm is a three-dimensional topological insulator. While HgTe does not exhibit an intrinsic bandgap in its bulk states, a strain-induced bandgap is opened [1,2] when a HgTe layer is epitaxially grown on a CdTe substrate. Owing to the topological properties of HgTe, a single family of gapless Dirac states appears at each of the two-dimensional surfaces [3]. Transport measurements [4] and terahertz experiments [5,6] have indeed confirmed that two-dimensional Dirac states exist at the surface of strained bulk HgTe. More recently, transport experiments on Josephson junctions based on these HgTe layers have demonstrated that superconductivity can be induced in the topological surface states by proximity to a superconductor [7]. Since Majorana fermions –particles which are indistinguishable from their own antiparticles– are expected to emerge at the interface between a superconductor and a topological insulator [8], superconducting weak links of strained bulk HgTe are promising solid state systems to investigate the manifestation of such Majorana quasiparticle excitations in transport experiments [9].

Josephson junctions are electronic devices consisting of two superconductors connected by a weak link. Owing to the coupling between both superconducting condensates, Cooper pairs can be transferred dissipationless from one superconductor to the other. This is the Josephson effect and gives rise to a supercurrent through a weak link, which has been studied extensively in a large variety of material systems, from thin insulating layers to conducting films, semiconducting nanowires, carbon nanotubes, and graphene [10–15]. When a conducting material is sandwiched between two superconductors, the proximity effect leads to induced superconductivity in the normal conductor, and electronic transport is mediated by Andreev bound states (which are phase-coherent superpositions of electron and hole excitations) [12,16]. It has been predicted that a topological insulator weak link supports non-chiral, zero-energy Andreev modes in the surface states (when the phase difference between both superconductors is $\phi = \pi$ or odd multiples of $\pi$) [8]. These zero-energy modes are a consequence of the induced p-like pairing symmetry of the topological surface states, and are intimately related to Majorana fermions, which give rise to a 4π-periodic Josephson effect [9,17].

Recent experiments have shown the first observation of a Josephson supercurrent through topological insulators $Bi_2Se_3$ [18,19] and $Bi_2Te_3$ [20,21]. However, these Bi-based materials exhibit a residual bulk conductance that dominates electronic transport, and obstructs an unambiguous investigation of induced superconductivity in the topological surface states. Contrary to the Bi-based topological insulators, the bulk of strained HgTe is effectively insulating [4], enabling the exploration of the Josephson effect in its topological surface states [7]. For this reason, we have fabricated lateral HgTe-based Josephson junctions (with closely spaced superconducting Nb electrodes on the top surface [22]) to study the occurrence of Andreev bound states at the surface, and to investigate superconducting transport through these unconventional states (Fig. 1a).

First we identify the origin of electronic transport in the 70 nm thick strained HgTe layer that we used as weak link. For this purpose, we have fabricated from the same wafer as the superconducting devices described below, a six-terminal Hall bar device with a channel length of 600 μm and width of 200 μm (same geometry as the device in Ref. 4), and measured the magnetic field dependence of its Hall resistance at cryogenic temperatures (Fig. 1b). From the low-field data, the mobility and charge density of the conduction electrons are extracted: $\mu \approx 26{,}000$ cm$^2$/Vs and $n \approx 5.5 \cdot 10^{11}$ cm$^{-2}$. Interestingly, figure 1b shows a series of quantization plateaus of the Hall resistance when the magnetic field is increased above $B \approx 2$ T. This observation provides evidence that transport is effectively through two-dimensional states. The magnetic field dependence of the Hall conductivity (inset of Fig. 1b) reveals the development of plateaus at $\sigma_{xy} = \frac{2e^2}{h}, \frac{3e^2}{h}, \frac{4e^2}{h}$ and weak plateau-like features close to $\sigma_{xy} = \frac{5e^2}{h}$ and $\sigma_{xy} = \frac{7e^2}{h}$ (comparable to the reported observations in Ref. 4). The appearance of quantum Hall plateaus corresponding to even as well as odd filling factors, with the odd ones as most robust at low magnetic fields, is characteristic of a single family of Dirac states at top as well as bottom surface with slightly different densities. This implies that the bulk is effectively free of mobile carriers, and electronic transport is predominantly through the two-dimensional surface states of strained HgTe [4–7].

Having identified that electronic transport occurs effectively through Dirac surface states, we can now investigate transport through a lateral Josephson junction, and study the nature of induced superconductivity in the topological surface states (Fig. 2a; this junction has been fabricated from the same HgTe layer by using two Nb contacts at the top surface as superconducting electrodes [22]). We measured the differential resistance $dV/dI$ of the junction as function of current bias, temperature and magnetic field in a measurement setup equipped with appropriate low-pass filters for supercurrent measurements [23]. In order to have the Nb electrodes in the superconducting state, the measurements were all done at temperatures well below the critical temperature of Nb ($T_c \approx 9$ K). Our measurements at the base temperature ($T = 25$ mK) of the dilution refrigerator clearly show two transport regimes (Fig. 2b): a dissipationless supercurrent appears in the junction when the bias is lower than a critical value $I_c$ (i.e., $R = 0$ if $I < I_c$), while a dissipating current occurs at higher current bias (with $R_n \approx 50\ \Omega$).

Let us inspect the resistive regime in more detail. When $dV/dI$ is plotted versus the measured voltage across the junction (Fig. 2c), we observe that the differential resistance slightly decreases when the voltage is below $V = 2\Delta/e \approx 2\ \mathrm{mV}$ (with $\Delta \approx 1\ \mathrm{meV}$ the superconducting gap of Nb), indicating that Andreev reflections occur at both HgTe-Nb interfaces, and that these interfaces are relatively transparent [24]. Moreover, a clear feature is present at $V = \Delta/e \approx 1\ \mathrm{mV}$, which can be attributed to multiple Andreev reflections [24]. The subharmonic gap structure related to higher order Andreev reflections is missing, probably because of smearing of the population of Andreev bound states in the course of multiple passages across a voltage-biased junction [25]. The smearing of the distribution of Andreev levels may be viewed as a nonequilibrium self-heating effect, as suggested previously for InAs-based superconducting junctions [26]. The inset of figure 2c shows that reproducible, aperiodic fluctuations are observable at $V < 0.4$ mV (where the current bias is just above the critical current). These fluctuations originate from the Josephson effect, and will be discussed in more detail later. The occurrence of multiple Andreev reflections at the HgTe-Nb interfaces gives rise to non-linear current-voltage ($I$-$V$) characteristics in the subgap regime (Fig. 2d; $V < 2\Delta/e$). Even for $V > 2\Delta/e$, Andreev reflection processes contribute to the total current through the junction [27]. This results in an excess current ($I_{\mathrm{exc}}$), which is usually of the order of the critical current, and can be determined from the measurements of Fig. 2d: $I_{exc} = I(V) - V / R_n \approx 4.5\,\mu A$ (where $R_n \approx 50\ \Omega$ is the differential resistance and $I(V)$ is the measured current at $V > 2\Delta / e$; see red dotted lines in Figs. 2b and 2d). Based on BTK theory [24,28] and using the values of $I_{exc}$ and $R_n$, the transparency of the HgTe-Nb interfaces can be estimated, yielding $T_{int} \approx 0.5$ [29]. This indicates that the interfaces have a moderate transparency [11].

When we inspect the supercurrent regime of the junction, we clearly observe hysteresis in the measured $I$-$V$ characteristics (Fig. 3a). The critical current corresponding to the transition from supercurrent to resistive regime is larger than the critical current corresponding to the transition from resistive to supercurrent regime –i.e., at $T = 25$ mK, the switching and retrapping current are $I_s \approx 3.8\ \mu A$ and $I_r \approx 2.5\ \mu A$, respectively. The temperature dependence of the $I$-$V$ characteristics shows that hysteresis is only present in the temperature range up to 1 K, whereas $I_s$ and $I_r$ exhibit approximately equal values at higher temperatures (Fig. 3b). Switching and retrapping current decrease with increasing temperature, until the supercurrent regime disappears at $T > 4$ K. Hysteresis usually occurs in underdamped Josephson junctions, i.e., in junctions that are effectively shunted by a large resistance and capacitance [10,11]. However, the capacitance of our lateral junction is very small $C \approx \varepsilon_0 t_{SC} W_{SC} / d \approx 34$ aF (where $\varepsilon_0$ is the vacuum permittivity, $t_{SC} \approx 90$ nm is the thickness of the superconducting electrodes and $W_{SC} \approx 8.5\,\mu m$ is the total width of the pair of parallel electrodes), and the Stewart-McCumber parameter of the RCSJ model is only $\beta_c = 2eI_c R_n^2 C / \hbar \approx 0.001$ [11]. Since $\beta_c << 1$, the junction is overdamped and no hysteresis is expected in the $I$-$V$ characteristics [11]. Alternatively, self-heating effects are known to lead to hysteresis in Josephson junctions with large critical currents [10,30]. In our junction, the power density at the retrapping point is estimated to be $P / V_{SS} \approx I_r V_r / V_{SS} \sim 10$ nW/$\mu m^3$ (where $V_{SS} \sim 0.01\ \mu m^3$ is an estimation of the volume corresponding to the surface states through which the current flows). This value indeed suggests that the observed hysteresis may well be due to electron heating (since it is consistent with Ref. [30]).

The $I_c R_n$ product –i.e., the critical current multiplied by the normal state resistance– is a characteristic junction parameter that provides useful information about superconducting transport through the Josephson junction. The $I_c R_n$ product is usually $\sim \Delta / e$ for short junctions (i.e., junctions with a Thouless energy $E_{Th}$ larger than the superconducting gap, $\Delta < E_{Th} = \hbar D / L^2$, where $D$ is the diffusion constant and $L$ is the junction length), while it is much smaller for long junctions (i.e., junctions with $E_{Th} < \Delta$) [31]. For our junction, we obtain $I_c R_n \approx 0.2$ mV at the lowest temperature, which is about five times smaller than $\Delta / e \approx 1$ mV. Possibly, this indicates that the junction is in the long junction limit, where the superconducting coherence length $\xi$ is smaller than the spacing $d$ between both Nb electrodes (i.e., $\xi < d \approx 200$ nm). Alternatively, the suppressed $I_c R_n$ product may be an indication of insufficient interface transparency [32].

The measurements discussed so far do not identify a contribution from zero-energy Majorana bound states to the supercurrent. Contrary to Andreev bound states in conventional materials, which give rise to a $2\pi$-periodic Josephson effect ($I_c \propto \sin \phi$), the presence of zero-energy Majorana states in topological insulators yields a $4\pi$ periodic Josephson effect ($I_c \propto \sin(\phi / 2)$) [8,9]. It is therefore beneficial to study the current-phase relation to identify a contribution from Majorana states to the supercurrent. The critical supercurrent is maximum at zero field, and decreases with increasing magnetic field in an oscillatory way, yielding a Fraunhofer diffraction pattern if the supercurrent flows uniformly through the junction [11]. The periodicity is $\Delta B = \Phi_0 / A$ for a $2\pi$-periodic supercurrent (where $\Phi_0 = h / 2e$ is the flux quantum, and $A$ is the junction area). However, it is predicted that the periodicity is twice as large for a $4\pi$-periodic supercurrent [9], i.e. $\Delta B = 2\Phi_0 / A$. Thus, exploring the magnetic field dependence of the critical current of our junction is expected to be an effective method to identify the nature of the supercurrent carrying states.

When we measure *I-V* characteristics as a function of an applied perpendicular magnetic field, we clearly observe a Fraunhofer-like diffraction pattern of the critical current, with a periodicity of $\Delta B \approx 1.1$ mT (figure 4a). The pattern deviates from a perfect Fraunhofer pattern, and indicates that the supercurrent is not fully uniform (note that measurements on different junctions have shown that this deviation is indeed sample-specific) [11]. To determine the magnetic flux through the junction, we need to consider the effective junction area, which is larger than the area between the electrodes, because the penetration length of the niobium films in a perpendicular magnetic field is approximately $\lambda \approx 350$ nm [7]. Thus, the periodicity of the Fraunhofer pattern corresponds very well to $\Delta B \approx \Phi_0 / A$, implying a dominating $2\pi$-periodic supercurrent through the topological surface states [33]. Note that we do not expect Josephson screening currents to affect our measurements, because the junction is in the narrow junction regime (i.e., the junction width is smaller than the Josephson penetration depth: $W < \lambda_J = \sqrt{\Phi_0 t_{HgTe} W / 2\pi \mu_0 I_c (d + 2\lambda)} \approx 3.3$ μm, where $\mu_0$ is the vacuum permeability and $t_{\mathrm{HgTe}}$ is the thickness of the HgTe layer). Also flux focusing effects are not expected to play a role in our junction, because the planar dimensions of the Nb electrodes are larger than $\lambda \approx 350$ nm (i.e., the Nb electrodes are in the wide-electrode limit [34]).

Besides a Fraunhofer pattern, figure 4a shows distinct features in the resistive regime at currents just above the critical current. These features are most pronounced above the main lobe of the Fraunhofer pattern, but are also clearly visible above the first few sidelobes. These features correspond to the *dV/dI* fluctuations that we already observed in Fig. 2c at small bias ($V < 0.4$ mV). To investigate the origin of these fluctuations, we have measured *dV/dI* versus $V$ at different magnetic field values ranging from $B = 0$ to 1.12 mT (Fig. 4b). These data show that the fluctuations are maximum at zero field, and decrease with increasing magnetic field. At $B \approx 1.1$ mT, when one flux quantum is captured in the junction ($B = \Phi_0 / A$), the fluctuations vanish completely. Moreover, figure 4a shows that the fluctuations appear and disappear with the same periodicity as the Fraunhofer pattern. Obviously, these *dV/dI* fluctuations are a manifestation of the Josephson effect in the resistive regime [10]. The exact origin of these fluctuations, which we observed in all HgTe junctions studied so far, remains however somewhat obscure.

In conclusion, our experimental data show that the supercurrent through the topological surface states of HgTe exhibits no clear signature of Majorana bound states. In contrast to some recent claims in the literature [19], this should not come as a surprise. Actually, theory expects that quasiparticle scattering in the junction removes unprotected, zero-energy modes, leaving an energy gap in the Andreev bound states –similar to conventional junctions [12,35]. Effectively, the $4\pi$-periodicity of the unprotected Andreev bound states turns into a $2\pi$-periodicity when the spectrum is gapped. Only the zero-energy Andreev bound states with the momentum perpendicular to the interface are topologically protected and remain ungapped, but since this number of modes is predicted to be orders of magnitude smaller than the number of unprotected modes [9,17], a $2\pi$-periodic Josephson effect may dominate the supercurrent –as is the case in our experiments. Moreover, the emergence of zero-energy Majorana states depends on the properties of the interface, such as the transparency and the position of the Fermi level with respect to the Dirac point of the topological states at the interface [36,37]. Although we do not observe a clear sign of the presence of protected Majorana bound states, our results show that strained bulk HgTe is a promising material system to get a better understanding of the Josephson effect in topological surface states, and to search for the manifestation of zero-energy Majorana states in transport experiments.

22. A HgTe layer of 70 nm thickness is grown on a CdTe substrate by molecular beam epitaxy. The HgTe layer is etched in an Ar plasma to get a 2 μm wide HgTe stripe. On the top surface of the HgTe stripe, two 70 nm thick Nb electrodes –with a spacing of 200 nm– are deposited by using e-beam lithography, ultrahigh vacuum sputtering, and lift-off techniques. Additionally, a protective bilayer consisting of 10 nm Al and 10 nm Ru, is deposited on top of the Nb electrodes. Note that, prior to Nb sputtering, the HgTe surface is cleaned by exposing it to a very mild, low-power Ar plasma. This results in more transparent HgTe-Nb interfaces without affecting the quality of the surface [7].

23. The transport measurements are performed in current-bias mode in an Oxford dilution refrigerator at a base temperature of 25 mK. The differential resistance is measured as function of current bias, temperature and magnetic field by using standard lock-in detection techniques, and the dc voltage across the junction is measured simultaneously (note that the data of Figs. 3a,b and 4a are obtained by pure dc measurements). The signals of all four wires to/from the junction are filtered at different temperature stages: low-pass filters at room temperature; RF "eccosorb" filters at 4 K; low-pass RC filters and Cu powder filters at 25 mK. Filtering is required to suppress the high-frequency noise, which is destructive for a Josephson supercurrent [13–15,18].

29. This value is the average transparency over all transport channels, and is estimated by using the general expression for the excess current in one-dimensional superconductor/normal metal/superconductor junctions, multiplied by the number of transport channels in the junction [24,28].

**ACKNOWLEDGEMENTS**

We thank E. M. Hankiewicz, B. Trauzettel and P. Brouwer for useful discussions. This work was financially supported by the German Research Foundation (DFG-JST joint research program "Topological Electronics" and Grant No. TK 60/1-1), the EU ERC-AG program (project 3-TOP), and the DARPA program.

**FIGURE CAPTIONS**

FIG. 1. (a) Schematic picture of a lateral Josephson junction based on a strained, undoped HgTe layer of 70 nm thickness. The Dirac states residing at the surfaces of the HgTe layer form the weak link between the Nb electrodes. Transport measurements on such a junction are performed in current-bias mode, while measuring the voltage across the junction to investigate superconducting transport through the surface states. (b) Transport measurements at $T \approx 70$ mK on a Hall bar device with $L = 600$ μm and $W = 200$ μm, based on the same HgTe layer as the Josephson junction. The magnetic field dependence of the Hall resistance $R_{xy}$ shows the development of plateaus, implying transport through two-dimensional states. The inset shows the quantization of the Hall conductivity $\sigma_{xy}$ at odd as well as even Landau level filling factors, characteristic of Dirac states at top and bottom surfaces with slightly different densities [4]. This demonstrates that the bulk conductance is negligible and transport is effectively through the topological surface states.

FIG. 2. (a) Geometric dimensions of the Josephson junction and the corresponding energy diagram showing Andreev modes at finite-energy ($\varepsilon > 0$) and zero-energy values ($\varepsilon = 0$). The presence of zero-energy modes is predicted to give rise to a $4\pi$-periodic supercurrent through the topological surface states [8]. However, normal scattering in the weak link can remove unprotected zero modes, leading to a $2\pi$-periodic Josephson effect [9,17,35]. (b) Differential resistance versus current bias at $T = 25$ mK (ascending bias sweep). In the high-current regime, the differential resistance converges to the value of the normal resistance $R_n \approx 50\ \Omega$. When the current is decreased below a critical value, the differential resistance drops to zero, corresponding to a supercurrent regime. (c) Differential resistance plotted versus the voltage across the junction at $T = 25$ mK. When $V < 2\Delta/e$, $dV/dI$ decreases due to Andreev reflections. The feature at $V \approx \Delta/e$ can be attributed to multiple Andreev reflections. The inset shows reproducible, aperiodic fluctuations, which are related to the Josephson effect. The red (blue) curve is a bias sweep in ascending (descending) direction. The red and blue dashed, vertical lines indicate the switching ($V_s$) and retrapping voltages ($V_r$), respectively. (d) Non-linear *I-V*

characteristic of the junction at $T = 25$ mK (ascending bias sweep). The red dotted lines with a slope corresponding to $R_n \approx 50$ Ω, cross the $V = 0$ axis at a finite current value. This is the excess current $I_{exc} \approx 4.5$ μA, which is directly related to Andreev reflections in the junction [24,25,27,28].

FIG. 3. (a) Ascending (red) and descending (blue) $I$-$V$ characteristics, showing pronounced hysteresis ($T = 25$ mK). The switching current, which is the critical current of the transition to the normal state regime, is larger than the retrapping current, which is the critical current of the transition to the supercurrent regime. (b) The temperature dependence of the switching and retrapping current ($I_s$ and $I_r$) shows that hysteresis is absent at $T > 1$ K. Note that the supercurrent regime disappears at $T > 4$ K. The inset shows hysteretic $I$-$V$ characteristics in the temperature range of 25–800 mK.

FIG. 4. (a) Colorplot of $dV/dI$ data versus current bias (ascending sweeps) and applied magnetic field ($T = 25$ mK). The $dV/dI$ data are numerically derived from measured $I$-$V$ characteristics at different magnetic field values. The yellow color corresponds to the supercurrent regime ($dV/dI = 0$) and the red color to the normal state regime ($dV/dI > 0$). A Fraunhofer pattern appears with a periodicity of $\Delta B \approx 1.1$ mT ($\approx \Phi_0/A$), providing evidence that the supercurrent is carried predominantly by $2\pi$-periodic Andreev bound states. The deviation from a perfect Fraunhofer pattern indicates a non-uniform supercurrent through the junction. (b) Measured $dV/dI$ versus the voltage across the junction (ascending bias sweeps) at different magnetic field values ($T = 25$ mK). The amplitude of the reproducible, aperiodic fluctuations is maximum at $B = 0$, decreases with magnetic field, and vanishes completely at $B \approx 1.1$ mT (when one flux quantum is captured in the effective junction area). This implies that the fluctuations originate from the Josephson effect. The effective junction length of 950 nm incorporates the magnetic penetration length of the Nb (see inset).

(a)

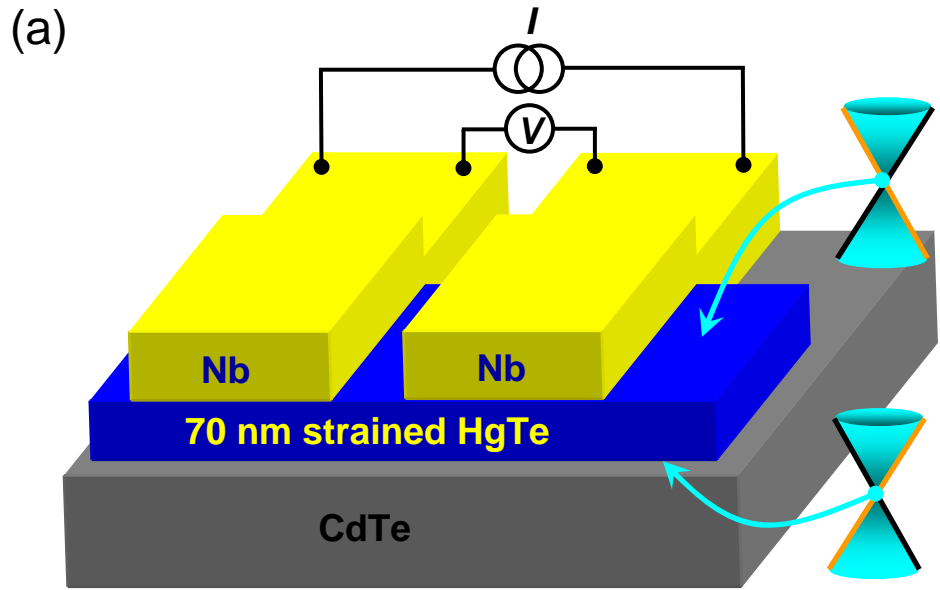

(b)

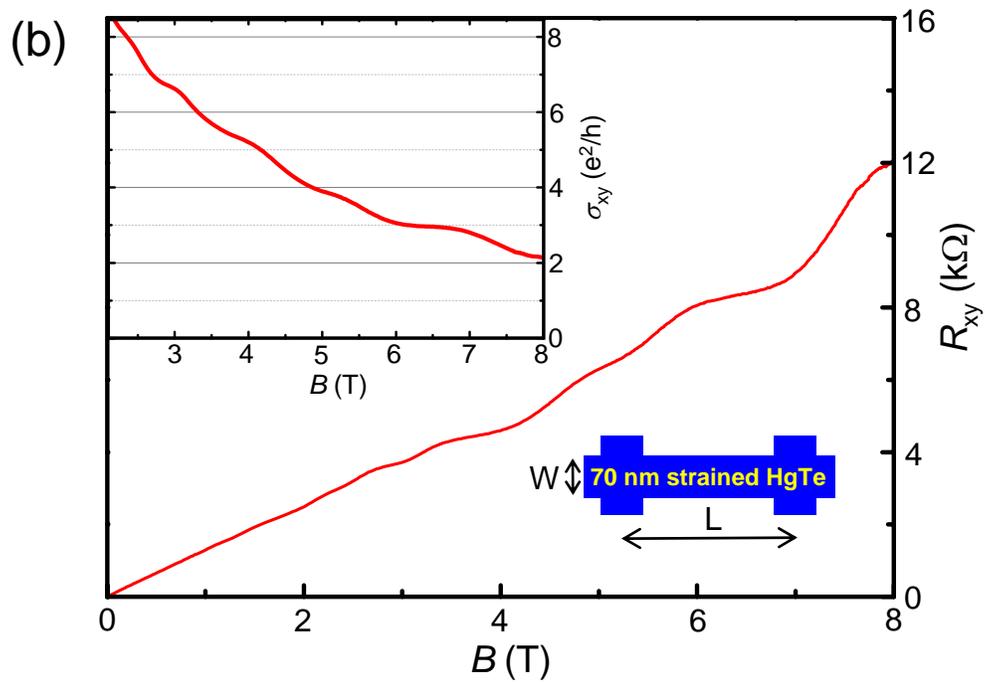

**Figure 2**

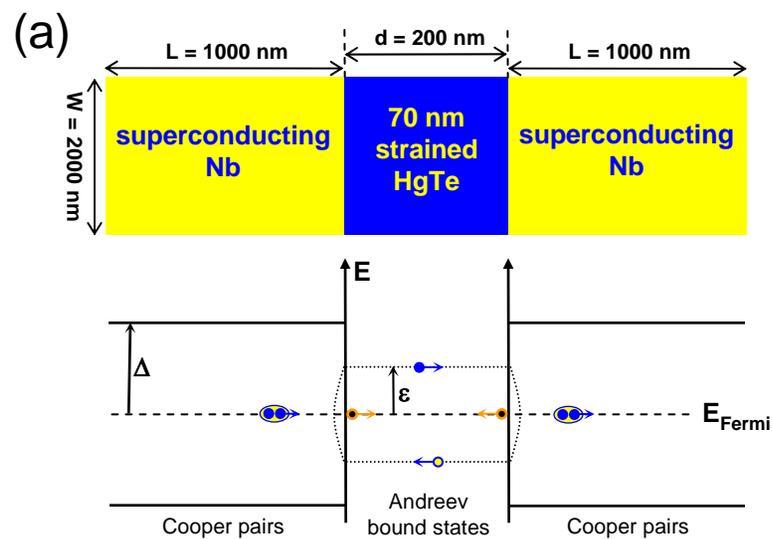

(a)

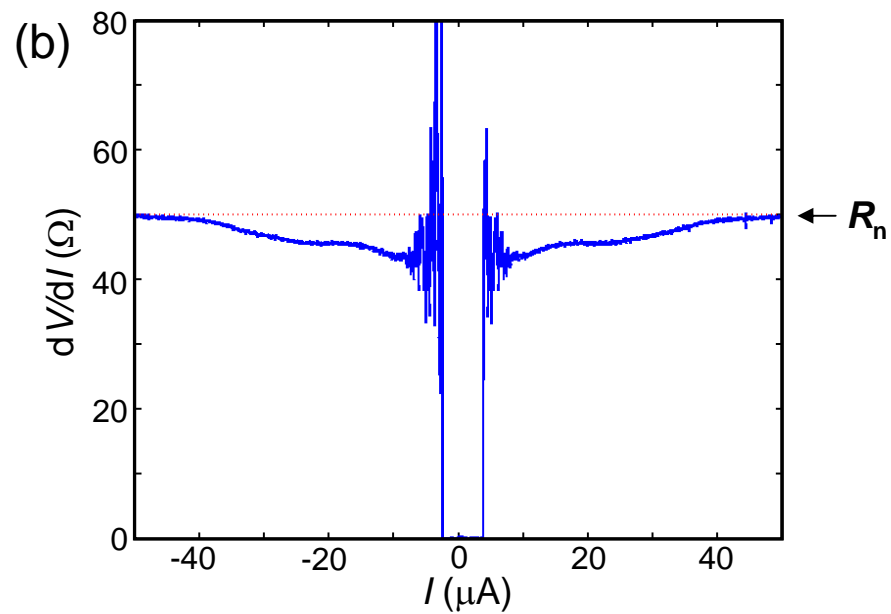

(b)

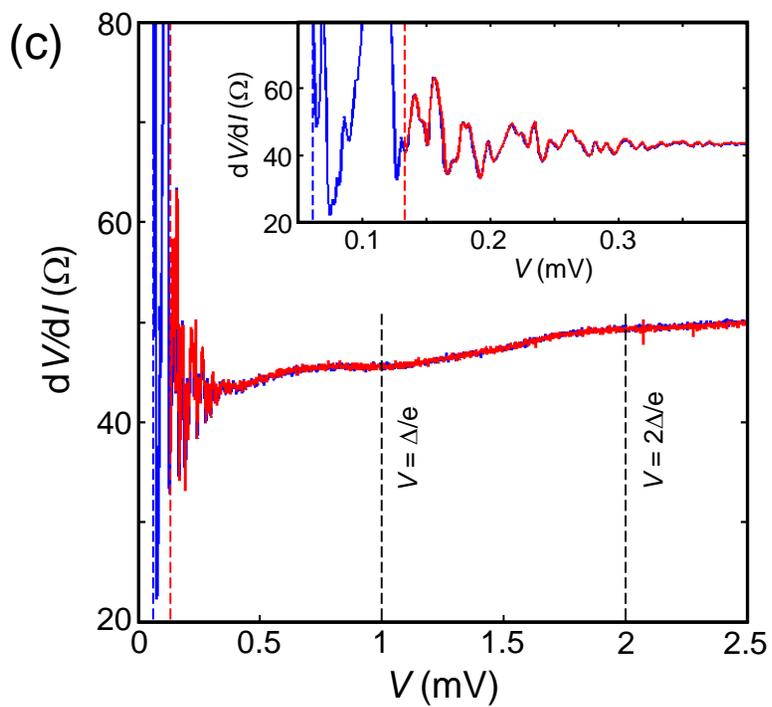

(c)

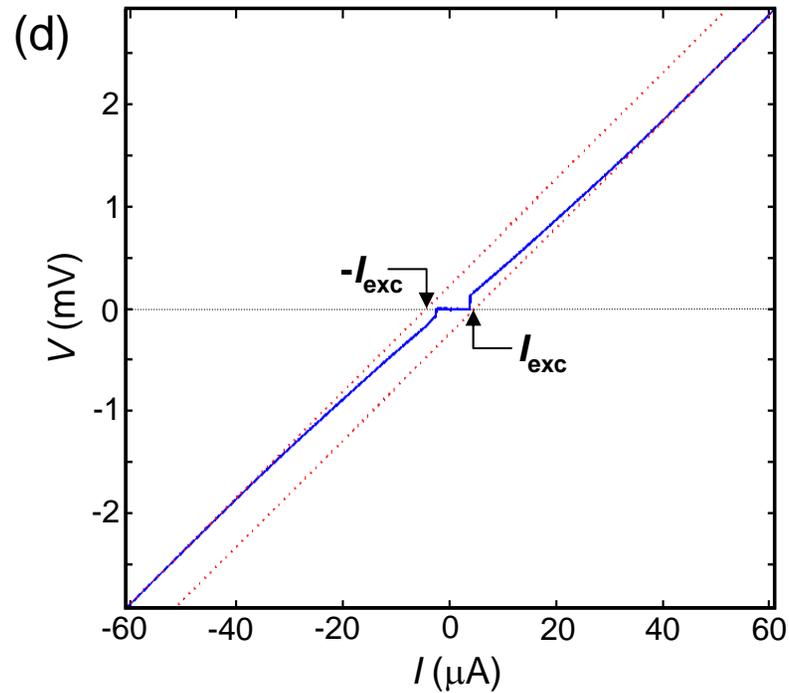

(d)

(a)

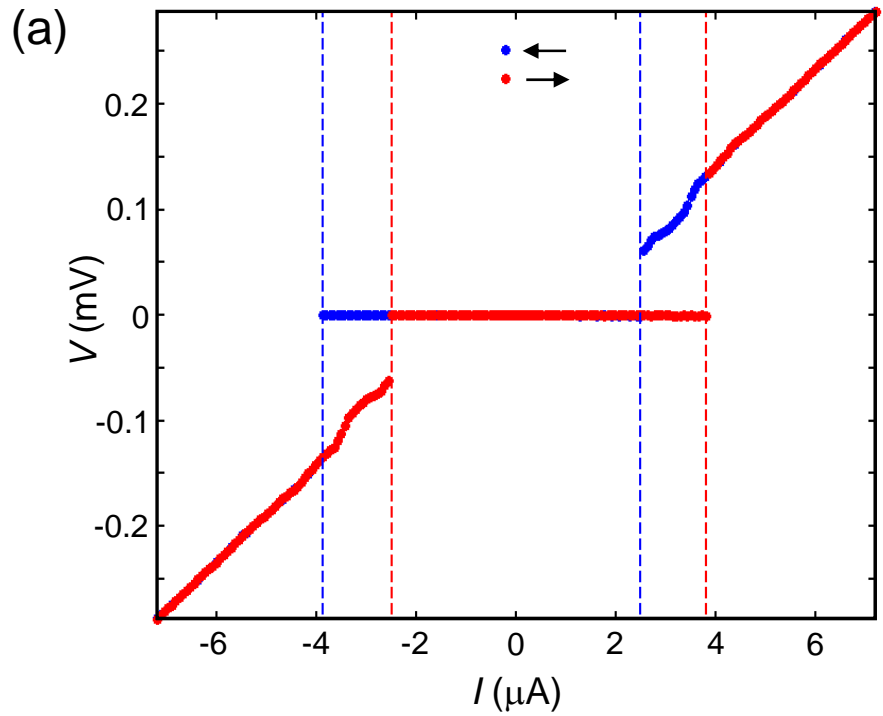

(b)

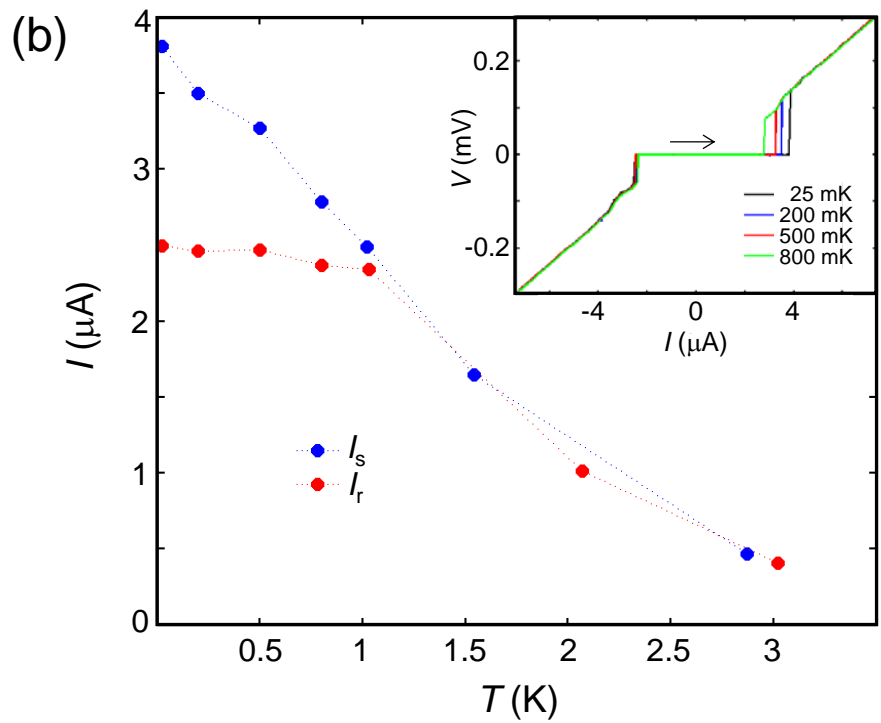

(a)

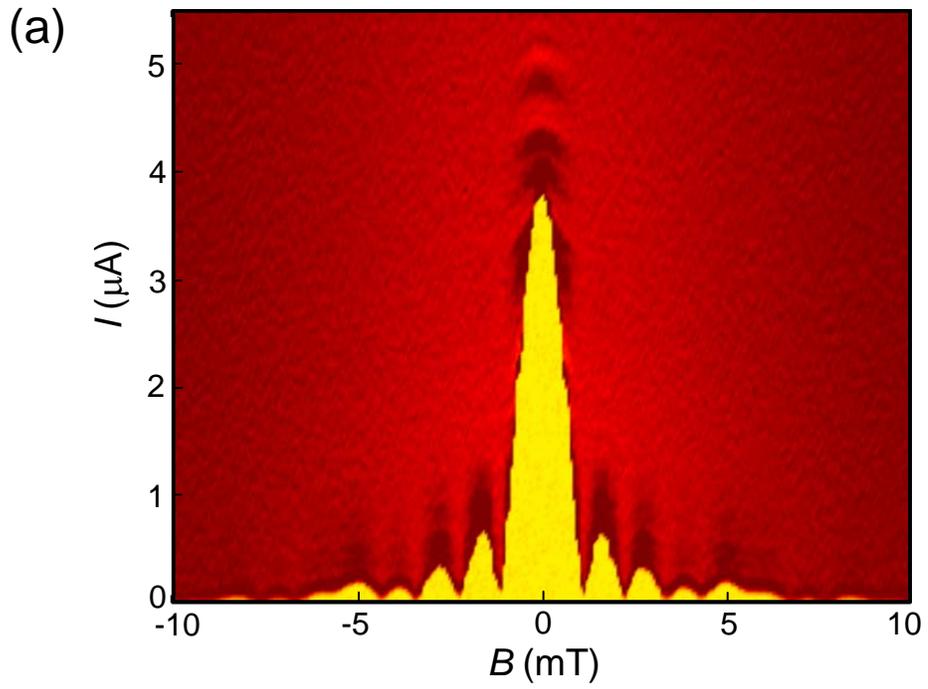

(b)

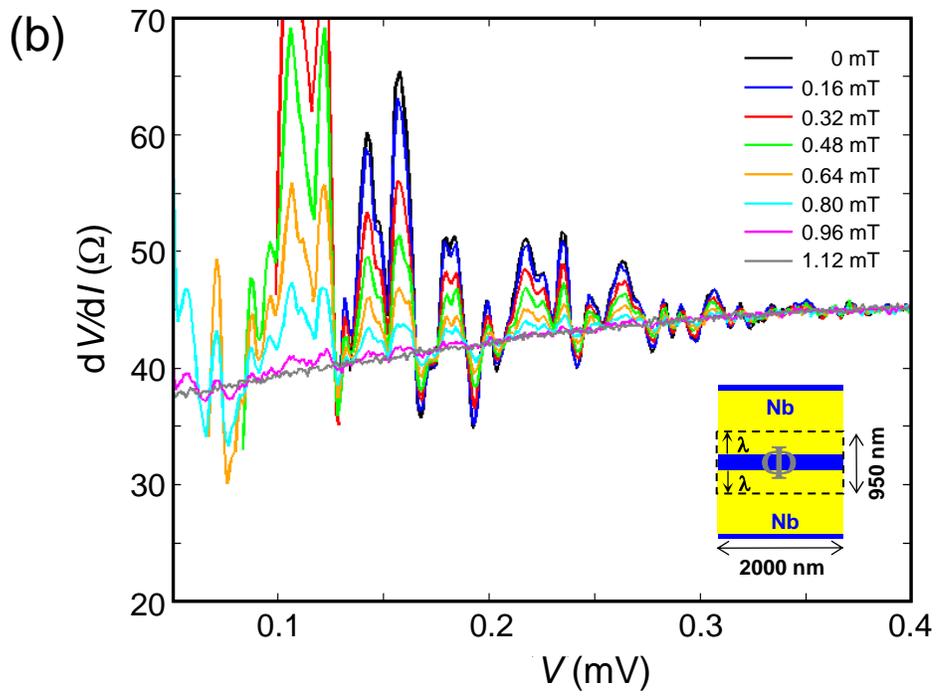